# Atomic layer deposition-based tuning of the pore size in mesoporous thin films studied by in situ grazing incidence small angle x-ray scattering

Jolien Dendooven,[*a] Kilian Devloo-Casier,[a] Matthias Ide,[b] Kathryn Grandfield,[†c] Mert Kurttepeli,[c] Karl F. Ludwig,[d] Sara Bals,[c] Pascal Van Der Voort[b] and Christophe Detavernier[a]

Atomic layer deposition (ALD) enables the conformal coating of porous materials, making the technique suitable for pore size tuning at the atomic level, e.g., for applications in catalysis, gas separation and sensing. It is, however, not straightforward to obtain information about the conformality of ALD coatings deposited in pores with diameters in the low mesoporous regime (< 10 nm). In this work, it is demonstrated that in situ synchrotron based grazing incidence small angle x-ray scattering (GISAXS) can provide valuable information on the change in density and internal surface area during ALD of $TiO_2$ in a porous titania film with small mesopores (3-8 nm). The results are shown to be in good agreement with in situ x-ray fluorescence data representing the evolution of the amount of Ti atoms deposited in the porous film. Analysis of both data sets indicates that the minimum pore diameter that can be achieved by ALD is determined by the size of the Ti-precursor molecule.

## Introduction

Atomic layer deposition (ALD) uses sequential self-terminating gas-solid reactions to deposit ultrathin coatings in a layer-by-layer fashion.[1] Its main advantages, atomic level control of the film thickness and excellent conformality on complex morphologies, render ALD a suitable technique for coating porous networks to create nanomaterials with improved compositional and/or structural properties.[2,3] These materials have applications in catalysis,[3-5] gas separation,[6-8] sensing,[9,10] dye-sensitized solar cells,[11,12] etc.

The implementation of ALD for the controlled tuning of pore sizes in porous materials requires characterization techniques that can evaluate the conformal nature of the deposited ALD layer. While ALD coatings deposited in pores with sizes > 10 nm are typically studied by electron microscopy techniques, this becomes challenging when applying ALD in smaller mesopores. Most of the reported studies concerning ALD in small mesopores aim for separation membrane applications and focus on the gas conductance and separation properties of the ALD modified membrane, rather than on the characterization of the ALD growth itself.[6-8] Only few studies have reported efforts to characterize ALD deposition in sub-10 nm pores. George and co-workers investigated ALD of $Al_2O_3$, $TiO_2$ and $SiO_2$ in ~5 nm tubular alumina membranes by deriving the pore diameter after each ALD half cycle from in situ $N_2$ conduction measurements.[13,14] This valuable in situ approach is, however, only applicable to membrane supports. To study the ALD modification of mesoporous thin films, some of the authors recently implemented ellipsometric porosimetry (EP) onto an ALD setup.[15] The in situ EP method can provide information on the change in accessible porosity and pore radius of a mesoporous film upon ALD deposition in its pores. A technical drawback of this technique is the need for heating and cooling in between subsequent ALD steps and EP measurements, as EP has to be performed at room temperature. This increases the duration of an in situ EP experiment during ALD. Another technique that has been introduced by the authors is synchrotron-based in situ x-ray fluorescence (XRF), a method which allows monitoring the amount of material that is deposited in a mesoporous thin film during each ALD cycle.[16] The conformal nature of the ALD coating in the mesopore network can be evaluated from this experiment by comparing the in situ XRF growth curve obtained during ALD on the mesoporous film with the reference curve for ALD on a planar substrate. Moreover, insights can be obtained on the initial growth behavior of ALD processes on the mesopore surface, which has proven valuable to study the effect of different plasma treatments on porous low dielectric constant (low-k) films.[17]

This work evaluates synchrotron-based grazing incidence small angle x-ray scattering (GISAXS) as an in situ technique to examine ALD growth in mesoporous thin films. In recent years, GISAXS has emerged as a powerful technique for the structural characterization of the pore arrangement in nanoporous thin films.[18-20] Here, it is shown that in situ GISAXS during ALD can be used to monitor the change in density and internal surface area with progressing growth, and hence provide information on the pore filling mechanism. The interpretation of the GISAXS data is discussed based on the results of complementary in situ XRF measurements and a detailed characterization of the original mesoporous film by EP and quantitative electron tomography.

## Experimental

### Synthesis and characterization of mesoporous titania thin films

As starting material, a ca. 80 nm thick mesoporous titania film was synthesized on a Si substrate via spin-coating of a Ti tetra-isopropoxide (TTIP) precursor solution that was prepared through a sol-gel method.[21] After aging, the titania film was calcined at 354 °C to remove the triblock copolymer-template (Pluronic F127).

The accessible porosity and pore size distribution of the as-grown mesoporous titania film were obtained from EP measurements conducted in a home-built vacuum chamber to which a spectroscopic ellipsometer (Woollam M2000U) and a system for dosing toluene vapor were connected.

A direct three-dimensional (3D) characterization of the mesoporous network was realized by quantitative electron tomography. First, a micro-pillar sample was prepared using a FEI Nova™ NanoLab 200 Dual-Beam SEM/FIB system. The micro-pillar was then mounted on a Fischione on-axis rotation tomography holder (model 2050).[22] A series of two-dimensional (2D) high-angle annular dark field scanning transmission electron microscopy (HAADF-STEM) micrographs were recorded over a tilt range of 180° with 2° tilt increments using a FEI Tecnai TEM operated at 200 kV. After alignment of the micrographs using a cross-correlation algorithm, the 3D volume was reconstructed using SIRT algorithm.[23] Visualization and pore size determination of the reconstructed series were performed with the Amira software.

### Atomic layer deposition of TiO$_2$ and synchrotron-based in situ characterization

The ALD depositions and in situ XRF and GISAXS measurements were carried out in the ultrahigh vacuum thin film growth facility installed at beamline X21 of the National Synchrotron Light Source (NSLS) at Brookhaven National Laboratory.[24] TiO$_2$ was deposited at 200 °C using tetrakis(dimethylamino)titanium (TDMAT, Ti(N(CH$_3$)$_2$)$_4$) as precursor and H$_2$O as oxygen source.[25] An ALD cycle consisted of 20 s precursor (with Ar carrier gas) exposure at ca. $10^{-3}$ mbar, 40 s pumping, 20 s H$_2$O exposure at ca. $10^{-3}$ mbar, and 5 s pumping. An x-ray reflectivity measurement after 160 ALD cycles on a planar SiO$_2$ substrate revealed a growth per cycle (GPC) of ca. 0.55 Å for this process.

Following each ALD cycle, a 50 s XRF measurement was performed using 10 keV (0.124 nm) photons at an incident angle of 5°. The spectrum was collected by a silicon drift detector mounted perpendicular to the sample surface. For an incident angle of 5°, the penetration depth of 10 keV x-rays in bulk TiO$_2$ is much larger than the thickness of the titania film under investigation, implying that the Ti Kα fluorescent signal is directly proportional to the number of Ti atoms in the film.

2D GISAXS scans were made every 2 ALD cycles using a Pilatus 100k detector (Dectris) mounted parallel to the sample at a distance of ca. 760 mm (Fig. 1). Using an integration time of 40 s, the scattering intensity was obtained as a function of the in-plane exit angle, $2\theta_f$, and the out-of-plane exit angle, $\alpha_f$. The incidence angle, $\alpha_i$, was 1°, which is well above the critical angle for total external reflection of bulk TiO$_2$ ($\alpha_c$ = 0.226° at 10 keV), meaning that the entire depth of the investigated titania film is probed. Another consequence is that the Yoneda peak, an enhancement of the scattered intensity at the exit angle which equals the critical angle of the sample (labelled Y in Fig. 1), and the specular peak (labelled S in Fig. 1) are well separated. Therefore, the intense specular reflected beam could be masked by a slit placed in between the sample and the detector without blocking the scattering pattern that contains information on the structure of the titania film.

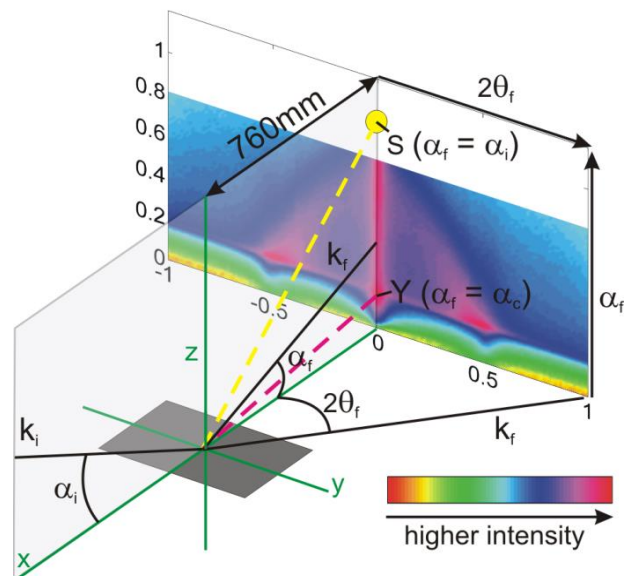

**Fig. 1** Schematic representation of the GISAXS geometry. The x-ray beam hits the sample at an incident angle $\alpha_i$ = 1°. The scattered beam is described by the exit angles $\alpha_f$ and $2\vartheta_f$ and the intensity is shown on a logarithmic scale. The specular reflected beam, S at $\alpha_f = \alpha_i$, is masked by a slit placed in between the sample and the detector. The Yoneda peak, Y, is observed at $\alpha_f = \alpha_c$.

## Results

First, the internal pore structure of the mesoporous titania film was studied in detail by TEM. Since the morphology of the pores was indiscernible from the 2D projections obtained by conventional TEM, it was necessary to perform electron tomography. This technique allowed us to reconstruct the 3D structure of the porous network from a series of 2D HAADF-STEM images acquired at successive tilts (Fig. 2(a)).[22,26] To examine the pore arrangement in more detail, various slices (the so-called "orthoslices") were characterized through the 3D reconstructed volume (Fig. 2(b) and (c)). While the in-plane orthoslices suggest the presence of spherical pores (Fig. 2(b)), the out-of-plane orthoslices clearly reveal the ellipsoidal shape of the pores (Fig. 2(c)). This high structural anisotropy is likely due to out-of-plane contraction of the porous network during calcination of the titania film.[27-29] The pore dimensions, the small and large axes of the ellipsoids, were extracted by manual analysis of ca. 200 randomly selected pores observed in different out-of-plane orthoslices. The results are presented in the histograms in Fig. 3. Average values of 3.8 ± 1.4 nm and 7.1 ± 1.6 nm were found for the pore small axis and large axis, respectively, yielding an anisotropy factor of roughly 2.

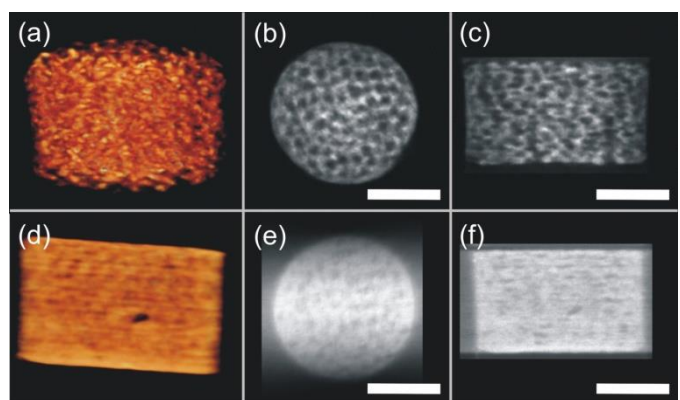

Fig. 2 Electron tomography study of the mesoporous titania film before (a-c) and after (d-f) modification via ALD. (a/d) 3D reconstructions of the pore networks. (b/e) In-plane and (c/f) out-of-plane orthoslices through the 3D reconstructions, where "white" = titania, "black" = pores. The scale bar indicates 50 nm.

The porosity and pore dimensions of the mesoporous titania substrate were further investigated by spectroscopic ellipsometry (SE) and EP. SE yielded a refractive index of 1.84 for the porous film, which corresponds to a total porosity of ca. 30% if a refractive index of 2.45 (anatase) is assumed for the titania matrix material. In the EP experiment, the ellipsometric angles $\Psi$ and $\Delta$ were recorded over the spectral range of 245-1000 nm during adsorption and desorption of toluene in and out of the porous film. The sorption isotherm was derived from the ellipsometric data according to a published method[15,30] and revealed an accessible porosity of ca. 15% (Fig. 4(a)). The shape of the hysteresis loop is associated with capillary condensation of toluene in ink-bottle shaped mesopores that are characterized by small pore interconnections (Fig. 4(b)).[31] The average size of these "bottlenecks" was ca. 3 nm, as deduced from the desorption branch using Kelvin's equation. A modified Kelvin equation was used for pore size determination from the adsorption branch to take into account the ellipsoidal shape of the pores.[27] This equation allows calculating the pore small axis size distribution for a known anisotropy factor, here assumed to be 2 based on the electron tomography study. The result is plotted in Fig. 3(a) together with a Gaussian fit to the data points. A good agreement was found between the EP data and the electron tomography study, suggesting that the inaccuracies inevitably introduced by manual analysis of the tomographic images were acceptably small.

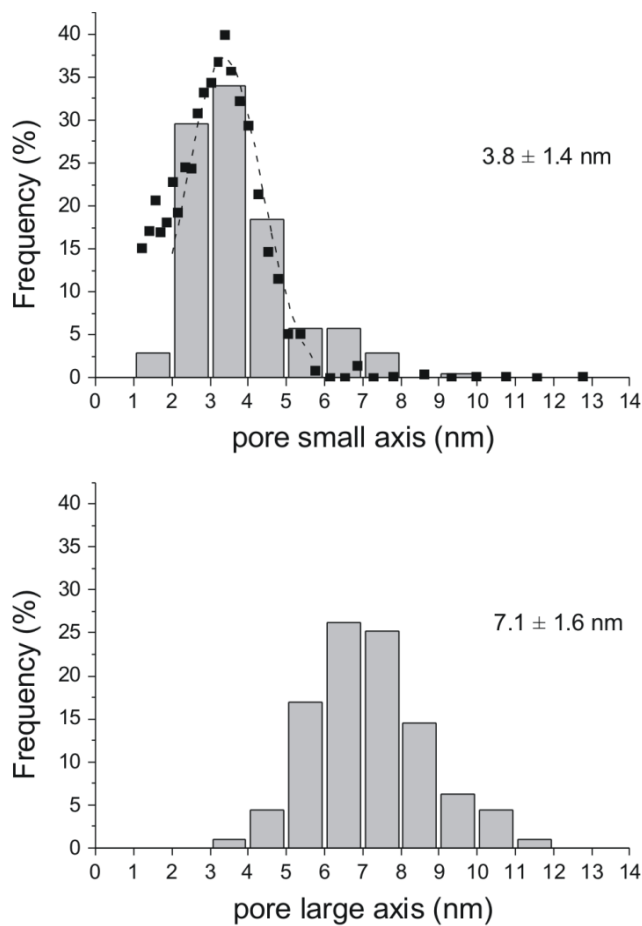

Fig. 3 The histograms represent the pore dimensions obtained from the electron tomography analysis. The average values are indicated in the graphs. The distribution of the pore small axis obtained from EP is plotted in the top graph (black squares). The dashed line is a Gaussian fit to the data.

Additional structural information was obtained from the 2D GISAXS pattern measured on the original titania film (Fig. 5). Besides the Yoneda peak and the off-specular reflectivity along $2\theta_f = 0$, an interference pattern is observed, which is indicative of spatial correlation between the pores in the film. The absence of clear scattering spots indicates, however, only short range order. To obtain the average distance between the pores, the detector plane described by $2\theta_f$ and $\alpha_f$ was converted to the reciprocal space, with $q_y$ parallel and $q_z$ perpendicular to the sample surface.[32,33] The elliptical shape of the scattering in $q$-space indicates that the average distance between the pores is different in the plane of the film than in the other directions, including the out-of-plane direction. An ellipse with center at the origin, major axis $a$, and minor axis $b$, was fitted to the interference maxima resulting in values for $a$ and $b$, from which the average out-of-plane and in-plane center-to-center pore distances were derived as $2\pi/a$ = 4.4 nm and $2\pi/b$ = 11.3 nm, respectively (Fig. 4(b)).[34] The GISAXS pattern thus confirmed the high structural anisotropy, in accordance to the electron tomography study.

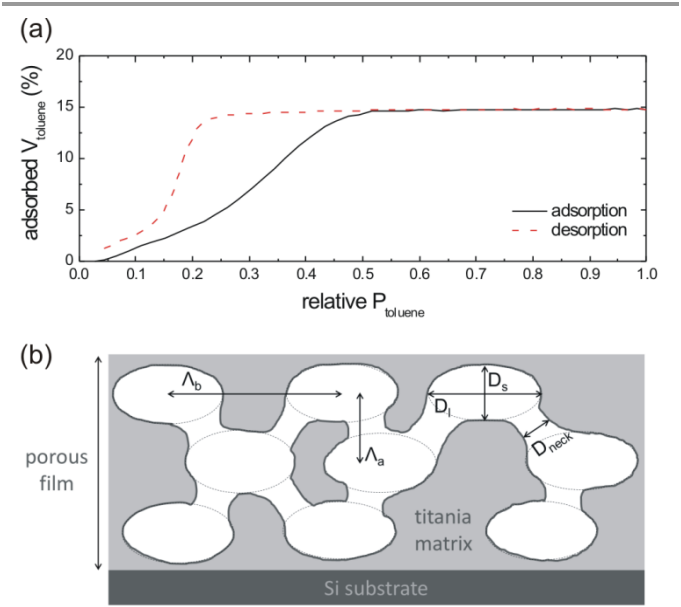

**Fig. 4** (a) Toluene sorption isotherm determined using EP on the mesoporous titania film. (b) Schematic representation of the pore arrangement in the mesoporous titania film with pore dimensions $D_s$, ~3.8 nm, $D_l$, ~7.1 nm, and $D_{neck}$, ~3 nm, as obtained from EP and electron tomography, and with pore distances $\Lambda_a$, ~4.4 nm, and $\Lambda_b$, ~11.3 nm, as obtained from GISAXS.

The mesoporous film was exposed to 40 cycles of the $TiO_2$ ALD process. This ALD treatment is expected to significantly decrease the pore size of the porous titania film by deposition of a conformal $TiO_2$ layer on the pore walls. The 3D reconstruction of the ALD modified film obtained with electron tomography is visualized in Fig. 2 (d-f). It is clear that the $TiO_2$ deposition has affected the internal pore structure of the original titania film, but the remaining tiny pores are not well defined and quantitative analysis is not possible for this data set. This result illustrates that characterizing ALD coatings in small mesopores by electron microscopy is challenging and that other methods are needed (Van Eyndhoven, G., Kurttepeli, M., Van Oers, C. J., Cool, P., Bals, S., Batenburg, K. J., & Sijbers, J. (2014). Pore REconstruction and Segmentation (PORES) method for improved porosity quantification of nanoporous materials. Ultramicroscopy, 1–10. doi:10.1016/j.ultramic.2014.08.008)

Here, we will demonstrate how valuable information can be obtained from in situ synchrotron-based GISAXS measurements. Although this technique is less direct than TEM, the in situ aspect offers the advantage to investigate the pore filling mechanism. Fig. 5 shows a selection of 2D GISAXS patterns recorded during the 40 cycles of $TiO_2$ ALD. Although the scattering pattern clearly changed with more $TiO_2$ deposited, ellipse fitting analysis revealed that the average out-of-plane and in-plane pore distances did not change, meaning that the pore arrangement is not affected by the ALD treatment, as expected. It is also clear that the Yoneda region, and thus the critical angle of the titania film, shifted to higher exit angles. To analyze this in detail, 1D out-of-plane GISAXS profiles were extracted at $2\theta_f = 0$ (vertical dotted lines in Fig. 5), and plotted in Fig. 6(a). The Yoneda peak position was determined from these profiles and the critical angle of the titania film was found to gradually increase from ca. 0.165° to ca. 0.195° during the first 24 ALD cycles (Fig. 6(b)). Because the square of the critical angle is proportional to the average electron density of the porous film, this observation provides evidence for the deposition of $TiO_2$ onto the interior surface of the mesopores, causing a gradual increase in density (electrons/cm$^3$) of the film. Starting from ALD cycle 24, the critical angle remained constant, meaning that the density of the porous film no longer changed, suggesting that the Ti-precursor molecules could no longer penetrate into the mesopores and that deposition continued on top of the coated mesoporous film.

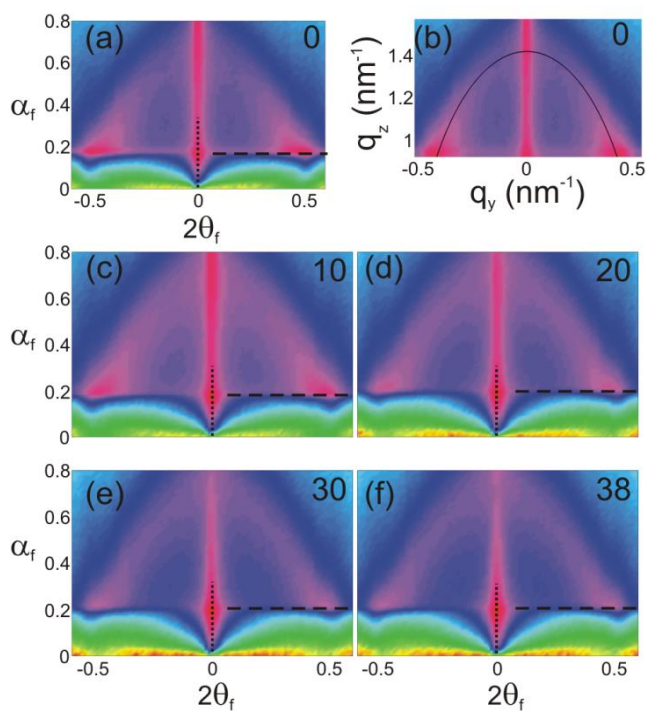

**Fig. 5** In situ GISAXS during ALD of $TiO_2$ in the mesoporous titania film. (a)/(c-f) 2D GISAXS patterns measured after 0, 10, 20, 30, 38 ALD cycles, respectively. The horizontal dashed lines indicate the Yoneda region. (b) 2D GISAXS pattern shown in (a) converted to the reciprocal space, with $q_y$ parallel and $q_z$ perpendicular to the sample surface.

Next, 1D in-plane GISAXS profiles were taken at the derived critical angles (horizontal dashed lines in Fig. 5), at which the information depth is of the order of a few tens of nanometers. Because the internal $TiO_2$/void interfaces are the actual scattering features, the scattering intensity can be interpreted as a measure for the internal surface area in the top layer of the porous titania thin film. By integrating the GISAXS profiles in Fig. 7(a), the scattering intensity can be plotted against the number of ALD cycles (Fig. 7(b)). During the first ca. 24 ALD cycles, the scattering intensity clearly decreased due to deposition of $TiO_2$ onto the mesopore walls, causing a gradual decrease of internal surface area. After ca. 24 ALD cycles, the scattered intensity further decreased, but much more slightly, due to the increasing thickness of the $TiO_2$ layer deposited on top of the mesoporous film.

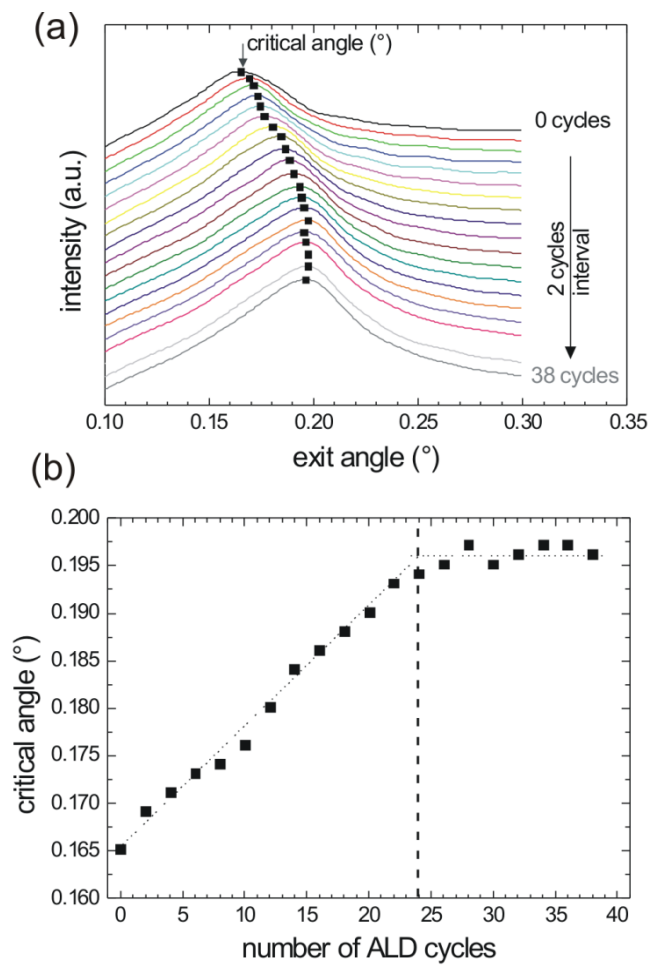

**Fig. 6** (a) Out-of-plane GISAXS profiles taken at $2\vartheta_f = 0$ (i.e. at the vertical dotted lines in Fig. 5). The individual curves are offset for clarity. (b) Critical angle, i.e. the position of the maximum of the interpolated curves shown in (a), against the number of ALD cycles. The dotted lines serve as guide to the eye.

The in situ GISAXS measurements were accompanied by XRF measurements to monitor the Ti-uptake during the deposition. After subtracting the initial Ti signal originating from the empty titania film, the Ti XRF intensity was plotted against the number of ALD cycles in Fig. 8. Comparison of the initial slope of the XRF intensity curve with the $TiO_2$ growth rate measured on a planar $SiO_2$ reference substrate revealed that ~9 times more Ti atoms were deposited in the mesoporous film. Assuming a conformal $TiO_2$ ALD coating (and a similar coverage of adsorption sites on the titania mesopore walls as on the planar $SiO_2$ surface), this result indicates that the accessible internal surface area of the original porous titania film is about 9 cm$^2$ per cm$^2$ of sample.[16] For comparison, a geometric surface area was calculated based on the accessible porosity and pore dimensions obtained from the EP measurement.[27] Values in the range 8-12 cm$^2$/cm$^2$ were found for on average 2-6 interconnections per pore. Since the surface area value derived from the XRF measurement falls within this range, it is reasonable to conclude that the growth of $TiO_2$ indeed proceeded conformally during the first ALD cycles. In agreement with the GISAXS data, the growth rate became similar to the reference value at cycle 23, indicating selective growth on the exterior surface of the coated film.

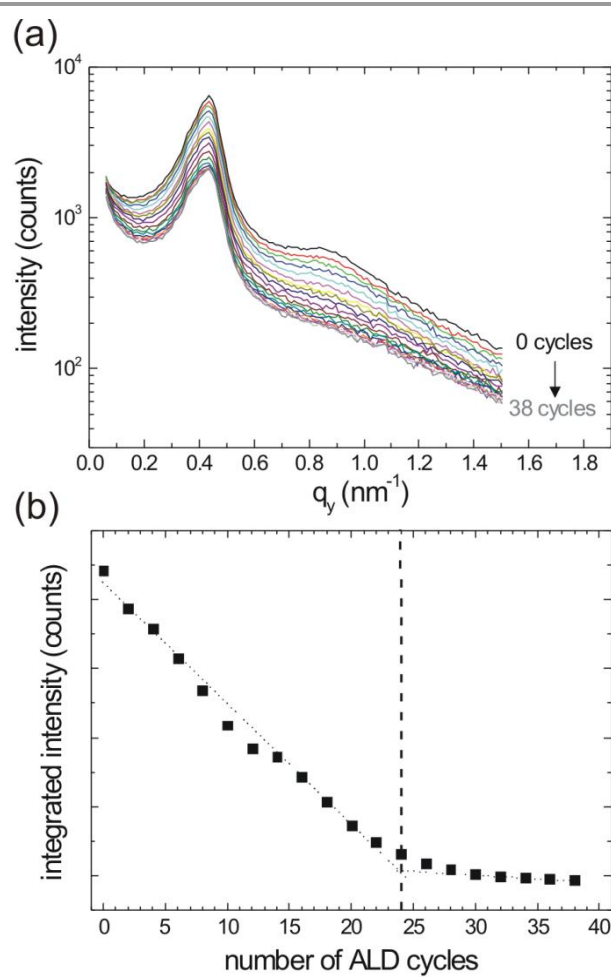

**Fig. 7** (a) In-plane GISAXS profiles at the respective critical angles (i.e. at the horizontal dashed lines in Fig. 5). (b) Integrated scattering intensity against the number of ALD cycles. The dotted lines serve as guide to the eye.

An SE measurement on the ALD modified mesoporous film yielded a refractive index value of 2.13, indicative of ca. 13% remaining porosity in the film. According to EP, this porosity was inaccessible for toluene vapour, in agreement with the interpretation of the x-ray data. The SE and EP data measured before and after the ALD treatment are summarized in Table 1. In principle, the total porosity can also be estimated from the critical angle measured in GISAXS.[36] Assuming a critical angle of 0.226° for the titania matrix material, the critical angles measured by GISAXS indicate an initial (and final) porosity of ca. 47% (and 26%) for the top layer of the uncoated (and coated) titania thin film. It should, however, be noted that an error of only one pixel in the conversion of the detector pixels to the scattering angle $\alpha_f$ would correspond to a systematic shift of all values in Fig. 7(b) by 0.013°, and a shift of ca. 10% on the calculated porosities. This is likely the main explanation for the discrepancy with the porosities calculated from the SE measurements.

GPC for $TiO_2$ ALD on a planar $SiO_2$ surface is 0.55 Å, and assuming a similar GPC in the titania ink-bottle mesopores, 23 ALD cycles are expected to result in a deposited $TiO_2$ thickness of 1.25 nm. Deposition of 1.25 nm $TiO_2$ on both sides of the 3 nm "bottleneck" will reduce the size of the pore interconnections to 0.5 nm, i.e. below the estimated kinetic diameter of the TDMAT molecule (ca. 7 Å). These calculations thus confirm that the pores were no longer accessible for the Ti-precursor molecules after 23 ALD cycles, while also confirming the feasibility of ALD to tune pore sizes down to the molecular level without clogging the pore mouths, as the x-ray scattering data provide a solid proof that TDMAT can still enter the pores up to cycle 23.

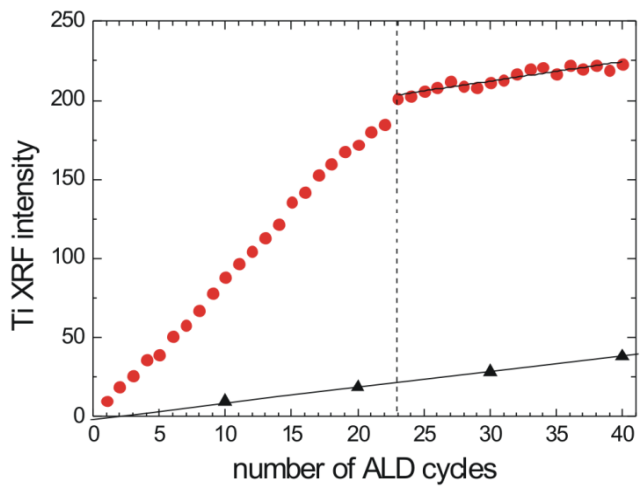

**Fig. 8** In situ XRF during ALD of $TiO_2$ on planar $SiO_2$ (black triangles) and in the mesoporous titania film (red circles): Ti Kα (4.5 keV) peak area against the number of ALD cycles. The solid lines are linear fits to the data points.

**Table 1** SE and EP characterization of the porous titania thin film before and after 40 ALD cycles of $TiO_2$

|  | Before ALD | After ALD |
| --- | --- | --- |
| Refractive index | 1.84 | 2.13 |
| Calculated total porosity (%)[a] | 30 | 13 |
| Accessible porosity (%)[b] | 15 | 0 |
| Pore small axis (nm) | ~3.8 | - |
| Pore large axis (nm) | ~7.1 | - |
| "Bottleneck" (nm) | ~3.0 | - |

[a] Assuming a refractive index of 2.45 (anatase) for the titania matrix material.
[b] Accessible for toluene having a kinetic diameter of 5.8 Å.[35]

## Discussion

The combination of electron tomography, SE, EP and GISAXS allowed us to get a detailed picture of the internal structure of the mesoporous titania substrate, resulting in the schematic representation in Fig. 9. During ALD of $TiO_2$ in the mesoporous film, both the in situ XRF and GISAXS measurements provided information on the pore filling mechanism and revealed a remarkably sharp transition at 23-24 ALD cycles. Both data sets indicated deposition inside the mesopores before this transition point, and deposition on top of the coated mesopores after the transition (Fig. 9). Given that the

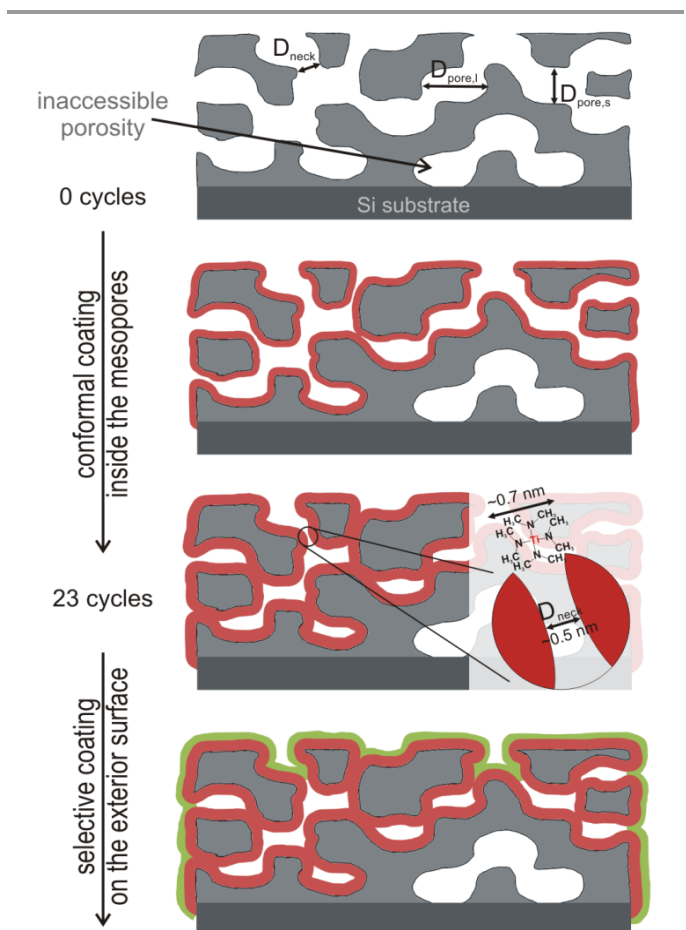

**Fig. 9** Schematic representation of the mesoporous titania film and the pore filling by ALD of $TiO_2$.

As can be seen in Fig. 5, the elliptical pattern is still present in GISAXS after the ALD treatment. This shows that there are still voids in the coated titania film, which is also confirmed by the SE measurement (Table 1) and the electron tomography study. Remaining porosity is not unexpected due to, on the one hand, the ink-bottle shape of the mesopores, and on the other hand, the presence of inaccessible voids, as observed from the significant difference between the total and accessible porosity in the original titania film (Table 1).

## Conclusions

The suitability of GISAXS to investigate the internal structure of mesoporous thin films was extended to study also ALD in this kind of materials. It was demonstrated that in situ GISAXS allows monitoring the evolution of the density and internal surface area during ALD of $TiO_2$ in a porous titania film. The results were in excellent agreement with in situ XRF data, and provided insights on the behavior of ALD in mesopores that were only accessible via 3 nm "bottlenecks".

## Acknowledgements

This research was funded by the European Research Council (ERC Starting Grant No. 239865 and No. 335078), by the Special Research Fund BOF of Ghent University (GOA 01G01513), and by the Research Foundation – Flanders (FWO). J.D. is a postdoctoral fellow of the FWO. K.L. acknowledges DOE for funding (Grant No. DE-FG02-03ER46037). Use of the National Synchrotron Light Source, Brookhaven National Laboratory, was supported by the U.S. Department of Energy, Office of Science, Office of Basic Energy Sciences, under Contract No. DE-AC02-98CH10886.

## Notes and references

[a] Department of Solid State Sciences, COCOON, Ghent University, Krijgslaan 281/S1, B-9000 Ghent, Belgium.
E-mail: Jolien.Dendooven@UGent.be
[b] Department of Inorganic and Physical Chemistry, COMOC, Ghent University, Krijgslaan 281/S3, B-9000 Ghent, Belgium.
[c] EMAT, University of Antwerp, Groenenborgerlaan 171, B-2020 Antwerp, Belgium.
[d] Physics Department, Boston University, 590 Commonwealth Avenue, Boston, Massachusetts 02215, United States.
[†] Present Address: Department of Materials Science and Engineering, McMaster University, Hamilton, Ontario, Canada.